\documentstyle[aps,multicol,prl]{revtex}
\begin{document}
\twocolumn
\draft
\title{Comment on ``Requirement of optical coherence for 
continuous--variable quantum teleportation'' by Terry Rudolph and 
Barry C. Sanders}
\author{H.M. Wiseman}
\address{School of Science, Griffith University, Nathan, Brisbane,
Queensland 4111, Australia}
\maketitle

\begin{abstract}
{\small The argument of Rudolph and Sanders, while technically correct, 
raises conceptual problems. In particular, 
if carried to its logical conclusion, it would disallow the 
use in our theories of any time $t$  with
implied resolution beyond that of direct human experience.} 
\end{abstract}  

\newcommand{\beq}{\begin{equation}} 
\newcommand{\eeq}{\end{equation}}
\newcommand{\bqa}{\begin{eqnarray}} 
\newcommand{\eqa}{\end{eqnarray}}
\newcommand{\nn}{\nonumber} 
\newcommand{\dg}{^\dagger}
\newcommand{\smallfrac}[2]{\mbox{$\frac{#1}{#2}$}}
\newcommand{\bra}[1]{\langle{#1}|} 
\newcommand{\ket}[1]{|{#1}\rangle}
\newcommand{\ip}[1]{\langle{#1}\rangle}
\newcommand{\sch}{Schr\"odinger } 
\newcommand{\schs}{Schr\"odinger's }
\newcommand{\hei}{Heisenberg } 
\newcommand{\heis}{Heisenberg's }
\newcommand{\half}{\smallfrac{1}{2}} 
\newcommand{\bl}{{\bigl(}}
\newcommand{\br}{{\bigr)}} 
\newcommand{\ito}{It\^o }
\newcommand{\str}{Stratonovich } 

In a recent paper \cite{RudSan01}, Rudolph and Sanders have 
argued that, contrary to  Ref.~\cite{Fur98}, 
continuous--variable quantum teleportation has 
not been and, in fact, cannot be, achieved using a laser as a source
of coherent radiation. They base their argument on the fact
that a laser is {\em not} a source of coherent radiation, in the 
sense that the output of a laser is not a coherent state, 
but a mixture of coherent states over all possible phases.
Although the formal analysis of Rudolph and Sanders is indisputably
correct, there are deep conceptual issues raised by this analysis 
that they seem not to have considered.

Having established, following M\o lmer \cite{Mol96}, 
the lack of absolute 
phase of a laser beam, Rudolph and Sanders say that they 
are not asserting that the production of coherent states of
light is impossible, and that  ``basic quantum electrodynamics
shows that a classical oscillating current can produce coherent
states.'' At first sight this seems unassailable, but let us
examine it more closely. 

First, how would one obtain a current
oscillating at optical frequencies? The natural oscillators
at optical frequencies are the electrons in atoms and molecules.
But how can one create coherent excitations in such oscillators if
one cannot start with coherent light? One answer would be 
to ``strike'' the atom (with a free electron, for example), to 
set it ``ringing''. Even assuming that the dynamics of the collision
can be fully determined, to produce a coherent oscillation,
the time of the collision would have to be known to an
accuracy less than an optical cycle, of order $10^{-15}$s.
 Otherwise one would have to average
over all possible phases and one would be left with exactly
the same problem as with the laser. 

Let us say for arguments sake that it is possible to know the 
time of collision to an accuracy of $10^{-15}$s. The question is,
with respect to what? What clock ticks $10^{15}$ times per second?
If we ignore that problem and allow Alice and Bob such clocks, 
ticking in phase, then surely this solves the problem? Well not really, 
because how do
we know that the clock really has a definite phase? How do we know
that, relative to the absolute time of the universe, Alice's 
 clock does not have a random phase? Certainly Alice cannot
simply look at her clock and verify that it does not have a random
phase, because she cannot perceive anything in $10^{-15}$s.

The point is that the whole idea of an absolute time standard
for the universe is highly questionable, even ignoring any issues to do with 
relativity. As conscious beings we feel that we experience time
directly, but experiments show that the limit of our time
resolution is in the range of tens or even hundreds of milliseconds 
\cite{Pen90}. 
On this basis, {\em it is 
impossible to establish the absolute phase of any oscillator
of frequency greater than a few tens of Hertz.} At higher frequencies 
we can only talk
about the phase of one oscillator relative to another oscillator. 
This conclusion is not altered by oscillations obtained by 
frequency $2^{n}$-upling, because the timing of the zeros of 
the highest harmonic 
can be no more accurately defined than that of 
the fundamental.

It follows then that,  if we accept the arguments of 
Rudolph and Sanders, we must conclude that it is impossible 
to teleport the state of a high-frequency oscillator by any means. 
Indeed, 
we should conclude that the state of any high frequency 
oscillator is always mixed (with all phases equally weighted)
so there is probably little point in trying to teleport it.

While this stance is a logically consistent one, it is far more
useful to acknowledge that there is no absolute time standard.
All we can ever do, for experiments involving time resolution beyond
direct human experience, is to use an agreed time standard. In this
context, a laser field ``ticking'' at $10^{15}$ Hz is as good a 
``clock'' as anything. {\em There are no better clocks, even in 
principle.}

To conclude, treating a laser ``clock'' as if it had a fixed phase relative to some 
absolute standard (that is, ascribing to it a coherent state)  
may be a ``convenient fiction'' \cite{Mol96}. It may even be committing the ``partition 
ensemble fallacy'' \cite{KokBra00}. However, the 
alternative, if carried to its logical conclusion, would be never to 
write down a time $t$ or a phase $\phi$ in our theories if its 
implied resolution would be beyond that of direct human experience. 
To scientists and engineers, this would be unacceptable pedantry.

\end{document}